 \documentclass[preprint]{elsarticle}
 \usepackage{graphicx}
 \usepackage{amsmath}
 \usepackage{amssymb}
 \usepackage{amsthm}
\usepackage[normalem]{ulem}
\usepackage{setspace}
\usepackage{txfonts}
\usepackage{subfigure}
\usepackage{fullpage}

\newcommand{\vect}[1]{\begin{pmatrix} #1\end{pmatrix}}

\newcommand{\dt}{h_n}

\begin{document}
\begin{frontmatter}
	\title{Exponential time integration for 3D compressible atmospheric models}

	\author[add1]{Greg Rainwater}\corref{mycorrespondingauthor}
	\ead{grainwater@outlook.com}
	\cortext[mycorrespondingauthor]{Corresponding author}
	
	\author[add2]{Kevin C. Viner}
	\ead{kevin.viner@nrlmry.navy.mil}
	\author[add2]{P. Alex Reinecke}
	\ead{alex.reinecke@nrlmry.navy.mil}
	
	\address[add1]{American Society for Engineering Education, 7 Grace Hopper Ave., Monterey, CA 93943}
	\address[add2]{U.S. Naval Research Laboratory,  7 Grace Hopper Ave., Monterey, CA 93943}

	\begin{abstract}
		 Recent advancements in evaluating matrix-exponential functions have opened	the doors to the practical use of exponential time-integration methods in numerical weather prediction (NWP). The success of exponential methods in shallow water simulations has led to the question of whether they can be beneficial in a 3D atmospheric model. In this paper, we take the first step forward by evaluating the behavior of exponential time-integration methods in the Navy's compressible deep-atmosphere nonhydrostatic global model (NEPTUNE-Navy Environmental Prediction sysTem Utilizing a Nonhydrostatic Engine).  Simulations are conducted on a set of idealized test cases designed to assess key features of a nonhydrostatic model and demonstrate that exponential integrators capture the desired large and small-scale traits, yielding results comparable to those found in the literature.  We propose a new upper boundary absorbing layer independent of  reference state and shown to be effective in both idealized and real-data simulations.  A real-data forecast using an exponential method with full physics is presented, providing a positive outlook for using exponential integrators for NWP.
		 
	\end{abstract}
	
\end{frontmatter}



\section{Introduction}
With the rapid and continuous evolution of atmospheric models, the development of efficient time integrators is a difficult and recurring problem.  The presence of fast acoustic and gravity waves rendered the use of explicit methods inadequate due to a severely restricted time step \cite{Ullrich2012}.  While an implicit scheme alleviates the allowable time step size restriction, these methods are computationally intensive and deemed infeasible without a suitable preconditioner. Techniques such as filtering approximations that eliminate the acoustic waves have been used to avoid the efficiency restrictions, but are not applied in this paper.  Even though such approximation was used in \cite{Robert1969,Robert1971}, it was their introduction of the semi-implicit methods that opened the doors to efficient time integration for numerical weather prediction (NWP) models.  By partitioning the dynamics such that the fast/slow waves can be handled implicitly/explicitly greatly improved the efficiency by reducing the stability constraints of the explicit methods and the complexity of the equations to be solved implicitly.  \cite{Klemp} proposed a similar technique known as split-explicit that does not rely on any manipulations of the equation set. The split-explicit methods uses a ``large" time step for slow dynamics and ``small" substeps to integrate the fast dynamics.  The relative simplicity and computational efficiency of the split-explicit methods made them an attractive and widely adopted time integrator for NWP \cite{Skamarock,wickerSkamarock,Klemp2007}.

Interest in implicit-explicit (IMEX) methods for NWP followed with the development of Runge-Kutta type IMEX methods \cite{ASCHER1997151,giraldoetal2013,Ullrich2012}. \cite{Ullrich2012} found that combining an explicit Runge-Kutta and linearly implicit Rosenbrock scheme provided time steps that were only restricted by the horizontal Courant-Friedrichs-Lewy (CFL) number.  \cite{gardneretal2018} thoroughly investigated the performance of twenty-one additive Runge-Kutta IMEX schemes for four different splitting strategies (or partitioning of the dynamics).  The results in \cite{gardneretal2018} indicate that Runge-Kutta IMEX schemes can offer a suitable option for global non-hydrostatic atmosphere simulations.

The aim of this article is to demonstrate the use of exponential integrators as an alternative time integration strategy for simulating nonhydrostatic compressible atmospheric dynamics.  Similarly to implicit methods, exponential integrators possess good stability properties, but require evaluation of exponential-like, rather than rational, matrix functions. Using a Krylov-projection-based method to evaluate an exponential-like function can save a significant amount of computational time compared to the rational function evaluation needed for an implicit integrator. Exponential methods have the additional advantage of being very accurate for nonlinear systems and exact for linear systems. The computational benefits of exponential integrators can be seen across a wide range of disciplines from computer graphics \cite{Michels2017} to magnetohydrodynamics \cite{EINKEMMER2017550}.  The use of exponential methods in NWP so far is limited, but positive results were reported for simulations of the shallow water equations on the sphere in  \cite{CLANCY2013665,GAUDREAULT2016827,LUAN2019817} .

The paper is organized as follows.  In Section \ref{sec:expint} we provide an overview of exponential integration and the integrators we will be using for our simulations.  Section 3 details our numerical experiments, where we evaluate the accuracy of the integrators and demonstrate their capabilities in simulating idealized test cases and real-data forecasts in the Navy's compressible deep-atmosphere nonhydrostatic model, NEPTUNE (Navy Environmental Prediction sysTem Utilizing a Nonhydrostatic Engine).

\section{Exponential Integrators}\label{sec:expint}
The construction of exponential integrators typically begins with reformulating 
\begin{equation}
u'(t)=f(u(t)),\quad u(t_0)=u_0, \quad u(t)\in \mathbb{R}^n
\label{eqn:ivp}
\end{equation}
as an integral equation which involves an exponential of either the full Jacobian $f'(u) \in \mathbb{R}^{n\times n}$ or a portion of it if a meaningful partitioning of $f(u)$ exists (e.g. if $f(u) = Lu+N(u)$ where $L$ and $N$ represent the linear and nonlinear part of $f(u)$ respectively).  As an example, consider linearizing $f(u(t))$ around $u_0$ using the first-order Taylor expansion such that \eqref{eqn:ivp} can be expressed as 
\begin{equation}
u'(t)=f(u_0)+f'(u_0)(u(t)-u_0)+r(u(t)),\quad u(t_0)=u_0, \quad u(t)\in \mathbb{R}^n,
\label{eqn:ivp_linearized}
\end{equation}
where $r(u(t))=f(u(t))-f(u_0)-f'(u_0)(u(t)-u_0)$.  Applying the variation of constants formula to \eqref{eqn:ivp_linearized} yields the integral form of the solution at time $t_0+h$:
\begin{equation}
u(t_0+h)=u_0+\left(\frac{e^{hf'(u_0)}-I}{hf'(u_0)}\right) h f(u_0)+\int_{t_0}^{t_0+h} e^{f'(u_0)(t_0+h-\tau)}r(u(\tau))d\tau.  \label{eqn:integral_form_ivp}
\end{equation}
Exponential methods are derived by applying various approximations to the integral term in \eqref{eqn:integral_form_ivp}.  The simplest approximation $r(u(t))\approx r(u_0) = 0$ produces the second-order exponential Euler (\textit{EXP2}) method
\begin{equation}
u_{n+1}=u_n+\left(\frac{e^{hf'(u_n)}-I}{hf'(u_n)}\right) h f(u_n),  \label{eqn:exp_euler}
\end{equation}
where $u_{n}\approx u(t_n)$.  
Higher-order methods are achieved by using different quadrature.  However, it can be seen from \eqref{eqn:exp_euler} that the numerical solution of an exponential integrator involves the  computation of an exponential-like function of a matrix.  This task was deemed computationally infeasible for large scale applications until Krylov projection-based methods were used to compute the exponential of a matrix and exponentially propagate the solution of a Hamiltonian problem from chemical physics \cite{nautswyatt}, and the Schr\"{o}dinger equation \cite{parklight}.  The extension to approximating a product of an arbitrary function of a matrix and a vector (by Van der Vorst \cite{vandervorst}) sparked interest in exponential integration with the development of many exponential propagation-based integrators \cite{cox, friesner, beylkin, gallopoulossaad,GAUDREAULT2021, hochbruckexp4, hochosterexpros, kassamtrefethen, krogstad,LUAN2017846,RainwaterTokman,rainwater2016,tokman2006}. 
The Exponential Propagation Iterative methods of Runge-Kutta type (EPIRK) \cite{tokman2006,Tokman} framework has been developed to allow derivation of efficient integrators particularly for large scale stiff equations that arise from large ODE systems or spatial discretization of PDEs.  The general formulation of an $s$-stage EPIRK method for the initial value problem (\ref{eqn:ivp}) is given by
\begin{eqnarray}
U_{ni} &=& u_n +a_{i1}\psi_{i1}(g_{i1}h_nA_{i1})h_nf(u_n)+\sum_{j=2}^{i-1} a_{ij}\psi_{ij}(g_{ij}h_nA_{ij})h_n\Delta^{(j-1)}r(u_n), \qquad i=2,\dots,s \label{eqn:epirk_internalstage}\\
u_{n+1} &=& u_n + b_1\psi_{(s+1)1}(g_{(s+1)1}h_nA_{(s+1)1})h_nf(u_n)+\sum_{j=2}^s b_j \psi_{(s+1)j}(g_{(s+1)j}h_nA_{(s+1)j})h_n\Delta^{(j-1)}r(u_n),\label{eqn:epirk_finalstage}
\end{eqnarray}
where $u_n$ is an approximation to the solution at some time $t_n$, $h_n= t_{n+1}-t_n$ is the time step and $\Delta^{(s)}$ denotes the forward difference operator
 constructed on stage-values $u_n,U_{n2},\dots,U_{ns}$ and is given by
 \begin{equation}
 \Delta^k r(u_n) \equiv \sum_{m=0}^{k-1} (-1)^m \begin{pmatrix} k \\ m\end{pmatrix} r(U_{n(k-m+1)}).
 \end{equation} 
  Different choices for functions $r(u)$, $\psi_{ij}$ and matrices $A_{ij}$ result in different classes of EPIRK methods \cite{Rainwater2017}.  According to our previous example using $r(u)$ as the first order Taylor expansion of $f(u)$, we take all matrices $A_{ij}=f'(u_n)$ to be the Jacobian evaluated at $u_n$ and functions $\psi_{ij}(z)=\sum_{k=1}^{p} p_{ijk}\varphi_k(z)$ are linear combinations of the exponential-like functions 
\begin{equation}
\label{eqn:PhiFunctions} \varphi_0(z)=e^{z}, \quad \varphi_k(z)=\int_{0}^1 e^{(1-\theta)z}\frac{\theta^{k-1}}{(k-1)!}d\theta, \quad k\geq 1.
\end{equation}
These choices result in a class of general (unpartitioned) EPIRK methods which are appropriate for cases where no meaningful partitioning exists for the operator $f(u)$ and the problem's stiffness comes from the full Jacobian matrix $f'(u_n)$.  Specific schemes can be constructed by deriving order conditions and solving to determine coefficients $a_{ij},b_{j}$, and $p_{ijk}$ (\cite{Tokman,tokman2006,rainwater2016}).  A closely-related class to the (unpartitioned) EPIRK schemes are the Exponential Rosenbrock (EXPRB) methods \cite{hochosterexpros,LUAN2014417}. In \cite{rainwater2016} it was shown that any EXPRB method can be written in EPIRK form.  By specifying the coefficients $a_{i1}=g_{i1}$ and functions $\psi_{i1}(z)=\varphi_1(z)$ in (\ref{eqn:epirk_internalstage}), it is also possible to express the EPIRK scheme in EXPRB form. 

The most efficient general exponential schemes result from coupling construction of particular integrators with the choice of an algorithm to approximate the matrix-function-vector products.  In general, when no useful information on the structure or the spectrum of the matrix is available or a matrix-free implementation is necessary then Krylov projection based algorithms are the methods of choice. A particularly efficient adaptive Krylov algorithm, \textit{phipm},  has been developed in \cite{niesenwright}. This method allows cost efficient computation of
\begin{itemize}
	\item[(I)] linear combinations of type $\varphi_0(\tau A)b_0 + \varphi_1(\tau A)b_1 + ... + \varphi_K(\tau A)b_K$ for a matrix $A$, vectors $b_i$, $i = 0,...,K$, and fixed $\tau$;
	\item[(II)] several terms of type $\varphi_k(\tau A)b_k$ involving a single function $\varphi_k$ and a vector $b_k$ for different values of $\tau$
\end{itemize}
Based on the ideas of \textit{phipm}, \cite{GAUDREAULT2018} introduced a new adaptive Krylov algorithm, \textit{KIOPS}, that significantly reduces the computational cost of computing (I) and (II).

The flexibility of the EPIRK \& EXPRB framework allows construction of exponential integrators that take advantage of the efficiency of the adaptive Krylov methods (\textit{phipm} and \textit{KIOPS}) and
reduce both the number and the cost of the matrix function vector products evaluations  \cite{GAUDREAULT2018,rainwater2016,LUAN201791}.  The exponential integrators we consider for our numerical experiments are a selection of such methods:
\begin{itemize}
	\item \textit{Exponential Euler (EXP2)} - stiffly accurate second-order integrator
	\begin{equation}
	\label{eqn:expEuler}
	\begin{array}{rl}
	u_{n+1}=& u_n + \varphi_1(h_nJ_n)h f(u_n) \end{array}\nonumber
	\end{equation}
	\item \textit{EPIRK4s3} \cite{Rainwater2017,Michels2017} - stiffly accurate fourth-order integrator
	\begin{equation}
	\label{eqn:epirk4s3}
	\begin{array}{rl}
	U_{n2} =& u_n  +  \frac{1}{8} \varphi_1(\frac{1}{8}h_n J_n)h_nf(u_n)                 \\
	U_{n3} =& u_n +  \frac{1}{9}\varphi_1(\frac{1}{9} h_n J_n)h_nf(u_n) \\
	u_{n+1}=& u_n + \varphi_1(h_nJ_n)h f(u_n) + \left(1892\varphi_3(h_nJ_n)-42336\varphi_4(h_nJ_n)\right)h_n\Delta r(u_{n})\\
	&\qquad \qquad + \left(1458\varphi_3(h_nJ_n)-34992\varphi_4(h_nJ_n)\right) h_n \Delta^2 r(u_{n}) \end{array}\nonumber
	\end{equation}
	
	\item \textit{EXPRB42} \cite{LUAN201791} - stiffly accurate fourth-order integrator
	\begin{equation}
	\label{eqn:exprb}
	\begin{array}{rl}
	U_{n2} =& u_n  +  \frac{3}{4} \varphi_1(\frac{3}{4}h_n J_n)h_nf(u_n)                 \\
	u_{n+1}=& u_n + \varphi_1(h_nJ_n)h f(u_n) + \frac{32}{9}\varphi_3(h_nJ_n)h_nr(U_{n2})\\
\end{array}\nonumber
	\end{equation}

\end{itemize}
The exponential Euler scheme requires only one evaluation of a matrix function vector product each time step, namely, $\varphi_1(h_nJ_n)h_nf(u_n)$.  Any higher-order scheme will require a minimum of two separate evaluations or calls to the adaptive Krylov algorithm.  \textit{EPIRK4s3} and \textit{EXPRB42} are examples of fourth-order methods that require only two Krylov projections per time step.  For \textit{EPIRK4s3}, the terms $\varphi_1(\frac{1}{8}h_n J_n)h_nf(u_n)$ and $\varphi_1(\frac{1}{9}h_n J_n)h_nf(u_n)$ are of type (II) and can be computed using a single call to the \textit{KIOPS} algorithm.  The second call is necessary for the evaluation of the terms (which are of type (I)) that appear in the final stage calculation.  The situation is similar for \textit{EXPRB42}, except that the first projection only computes one term ($\varphi_1(\frac{3}{4}h_n J_n)h_nf(u_n) $).  While efficient fifth-order methods have been constructed (see \cite{rainwater2016,LUAN2014417}), they require three or more Krylov projections per time step; since the major computational cost of exponential integrators is the evaluation of these matrix function vector products, it is unreasonable to consider these methods here.

\section{Numerical Experiments}
The purpose of our numerical experiments is to evaluate the behavior of exponential methods in a nonhydrostatic compressible global model.  All of the experiments presented below were conducted in the 3D spectral-element based atmospheric model, NEPTUNE, which is a compressible deep-atmosphere nonhydrostatic global model based on the equation set Set2NC in \cite{giraldoetal2010}. Starting with a fundamental set of idealized test cases, we assess the accuracy and the ability to resolve both the fast and slow processes.  We can confirm exponential methods produce comparable results using well-documented simulations in the literature.  We conclude by showing reasonable real-data forecasts can be obtained using exponential integrators.  

\subsection{Order verification and comparative performance}
To evaluate the accuracy of the time integrators, we apply them in NEPTUNE on the gravity wave test case defined in \cite{ullrichetal2012}.  This test case adds a potential temperature perturbation to an initially balanced atmosphere on a reduced Earth-radius sphere (1/125) that triggers the development of gravity waves.  Our spatial domain was discretized using 20 elements per edge of the cubed sphere, fourth-order spectral element polynomials, and 13 vertical levels (3 vertical elements) with a model top of $10~ km$. This corresponds to approximately $1.5^\circ$ ($1.3~ km$) horizontal grid spacing and $0.77~km$ in the vertical.  For each integrator, this test case is simulated for 1 hour using time steps of 1, 5, 10, and 15 seconds.  A reference solution was computed using a fully explicit third-order five-stage strong-stability-preserving Runge-Kutta method  (SSPRK(5,3)) \cite{RuuthRKSSP} using a time step of $0.0001$ seconds.  

\begin{figure}[ht]%
	\centering
	\subfigure[][ Log-log plots of error vs. time step size.  The dash-dot/dashed lines are lines with slopes equal to two/four respectively.]{%
		\label{fig:order}%
		\includegraphics[width=0.45\linewidth]{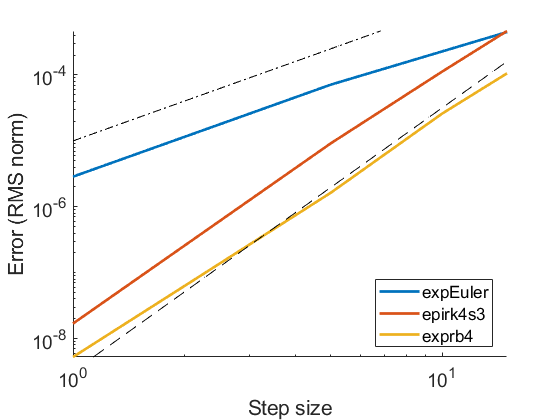}	}%
	\hspace{8pt}%
	\subfigure[][Comparative precision diagrams (CPU execution time versus error)]{%
		\label{fig:precision}%
		\includegraphics[width=0.45\linewidth]{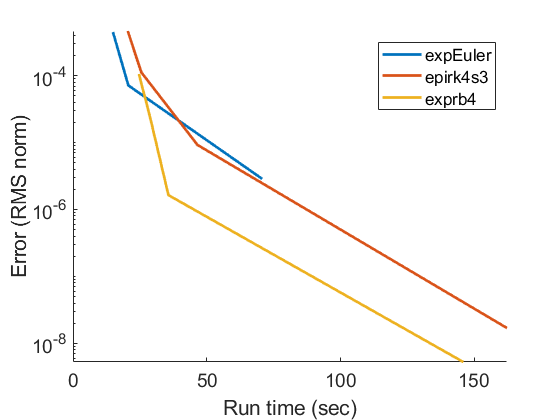}}\\		
	\caption[ Order and precision diagrams for the gravity wave test case from \cite{ullrichetal2012}]{ Order and precision diagrams for the gravity wave test case from \cite{ullrichetal2012} }
	\label{fig:orderEffdiagram}
\end{figure}

Figure \ref{fig:order} displays the plots of the time step versus error for the exponential methods outlined above.  For convenience, lines of slope two and four are plotted (dash-dot/dashed, respectively).  We can see that \textit{EXP2} and \textit{EPIRK4s3} achieve their full predicted order for the smaller time step sizes while \textit{EXPRB42} shows slightly less than fourth-order.  Considering all the time steps, the estimated order of convergence for the exponential methods are \textit{EXP2} $\sim 1.86$, \textit{EPIRK4s3} $\sim 3.77$, and \textit{EXPRB42}  $\sim 3.65$.

The comparative performance of the exponential integrators is shown in Figure \ref{fig:orderEffdiagram}(b).  Since the exponential Euler method requires fewer Krylov projections per time step, it was expected to be the most efficient for any given step size. Without any regard for the accuracy, the most efficient time step for each integrator is $h=15$ seconds.  Comparing the execution times for this step size, \textit{EXP2} is found to be $25\%$ / $38\%$ faster than \textit{EPIRK4s3}/\textit{EXPRB4}, respectively.  These values increase to $65\% $ / $50\%$ if we consider the execution times of \textit{EPIRK4s3} and \textit{EXPRB4} at their intersections with the graph of \textit{EXP2} (i.e., the points at which the fourth-order methods can be of benefit)  and compare them to the execution time of \textit{EXP2} using a time step of $h=15$ seconds.  

On the other hand, for a prescribed accuracy, the higher-order methods can offer significant computational savings.  For example, if we consider a error tolerance of $5\times 10^{-6}$ in Figure \ref{fig:orderEffdiagram}(b), then \textit{EXPRB4} is a factor of two faster than \textit{EXP2}.  The reason for this dramatic difference is that the lower order methods are forced to take a smaller step size which reduces the number of Krylov vectors per time step, but the computational cost becomes dominated by the large number of steps and the total number of Krylov vectors needed for the entire integration.  The accuracy of the fourth-order exponential methods (of Runge-Kutta type) comes from taking smaller step sizes; sacrificing the competitive edge of these methods.  While exponential multistep type methods \cite{cox,tokman2006,hockbruckOstermannExpMultiStep} are not included here, the recent development of high-order exponential multistep schemes \cite{GAUDREAULT2021} is the next step forward in achieving high accuracy (up to order seven) at near the computational cost of \textit{EXP2}.

\subsection{Mountain-wave simulations}\label{subsec:mtwave}

In this section we verify the ability of exponential methods to accurately capture nonhydrostatic processes by considering idealized mountain-wave test cases from \cite{Klemp2015}.  These simulations are on a reduced Earth-radius sphere (1/166.7) without rotation and designed to help evaluate the behavior of nonhydrostatic global models.  
\cite{Klemp2015} showed that simulations performed with the Model for Prediction Across Scales (MPAS) were in close agreement with analytical solutions of similar test case(s) in a Cartesian geometry \cite{schar2002}.  The results in \cite{Klemp2015} will serve as baseline for our comparisons.

The mountain-wave simulations from \cite{Klemp2015} put to use an implicit Rayleigh damping of the vertical velocity \cite{klempetal2008}. The idea of \cite{klempetal2008} (with the focus on time-split integration) is to make an adjustment (on the small time steps) to the vertical velocity such that it produces a Raleigh damping term and a (implicit) vertical diffusion term in the vertical momentum equation.  Unlike the time-split integrators, exponential methods  handle the fast acoustic and gravity waves via exponential-like functions and thus not able to directly apply the damping scheme from \cite{klempetal2008}.  Using the same idea, we propose an effective absorbing layer via an adjustment to the vertical velocity at the end of the time step for exponential integrators. As an example, the adjustment combined with the \textit{EXP2} method is
\begin{equation}\label{eqn:EulerAdjustment}
\left\{\begin{array}{l}
\widehat{\mathbf{u}}_{n+1}=\mathbf{u}_n+\dt \varphi_1(\dt J_n)f(\mathbf{u}_n)\\
\mathbf{u}_{n+1}=\mathbf{Q}\widehat{\mathbf{u}}_{n+1}
\end{array}\right.
\end{equation}
where $\mathbf{u}=(\rho,u,v,w,\theta)^T$ is the vector with density, velocities, and potential temperature as components, $\mathbf{Q}=\textrm{diag}\left[1,1,1,(1+\dt R_w)^{-1},1\right]$ and $R_w$ is the damping profile.  For our numerical experiments, we use the same damping profile as that used in \cite{Klemp,klempetal2008,Klemp2015}:
\begin{equation}
\label{eqn:dampingProfile}
R_w(z)=\alpha \sin^2\left(\frac{\pi}{2}\frac{z-z_h}{z_t-z_h}\right),
\end{equation}
where $z_t$ denotes the top of the models domain and $z_h$ is the bottom of the absorbing layer.  The adjustment to the vertical velocity in \eqref{eqn:EulerAdjustment} filters into the other equations on the next time step calculation which produces the desired Raleigh damping term and a vertical diffusion term in the vertical momentum equation.  A derivation of the exact terms is provided in \ref{sec:appendixAnalysis}.

\subsubsection{Quasi Two-Dimensional Mountain Ridge Case}
\begin{figure}[ht]%
	\centering
	\includegraphics[width=0.75\linewidth]{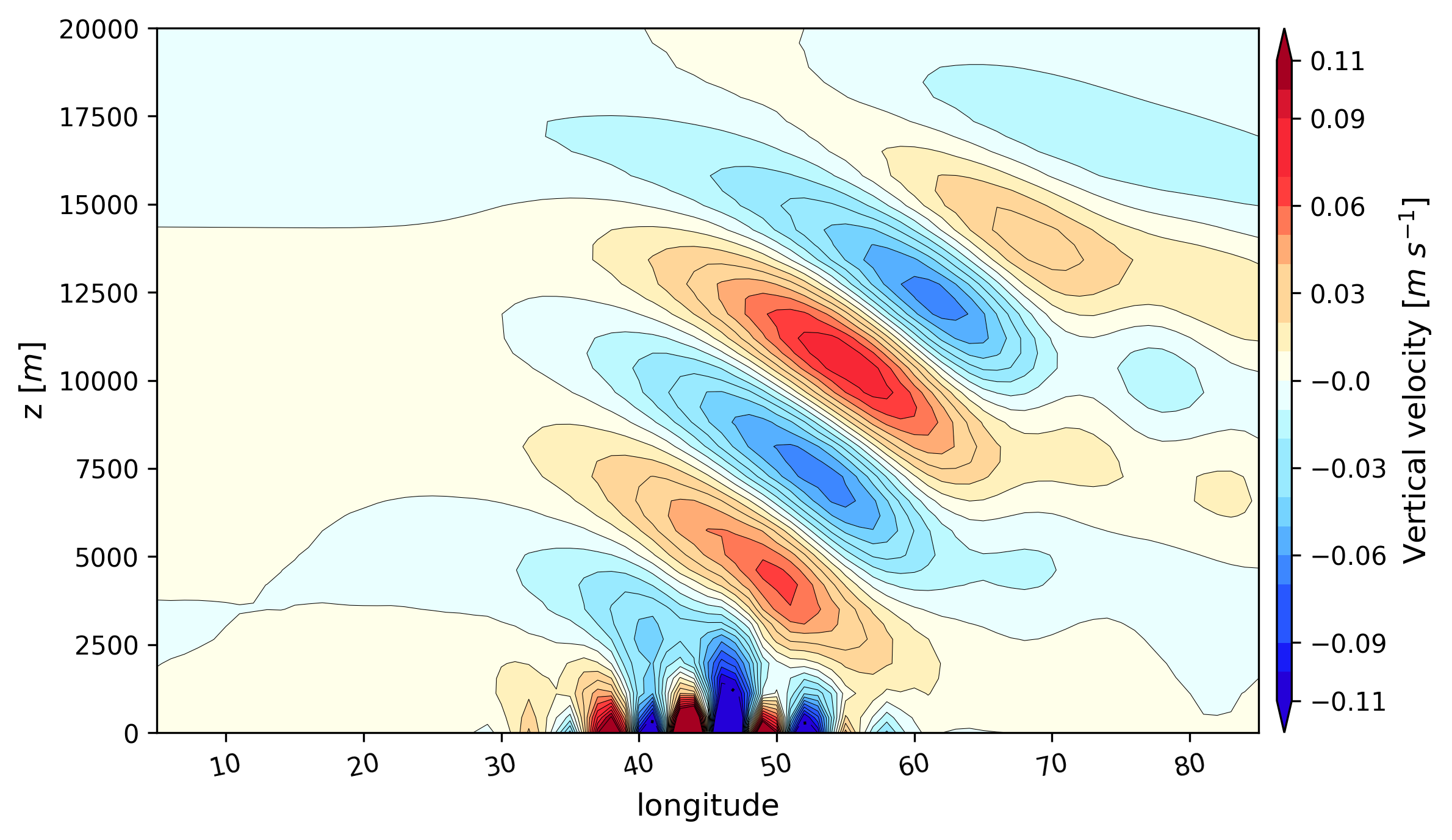}
	\caption{Vertical velocity along equator after 120 minutes simulation of the mountain ridge case. The vertical velocity contour interval is $0.01$ $m/s$.  }
	\label{fig:MtnRidge}
\end{figure}

This test case simulates the flow of an atmosphere with constant velocity over a specified mountain ridge terrain profile.  The details of the terrain profile (mountain ridge) can be found in \cite{Klemp2015}. 
For comparison purposes, we configure our model to be as close as possible to MPAS.  Thus for this test case, we specify a model top of $z_{\textrm{top}}=20~km$, with an upper boundary absorbing layer of depth $z_{\textrm{h}}=10 ~km$.  We used 28 elements per face of the cubed sphere, 3rd order polynomials, and 40 vertical levels (13 elements) to obtain approximately 1.1$^\circ$ ($720~m$) horizontal grid spacing and $500~m$ in the vertical. The exponential Euler (\textit{EXP2}) method was used to complete the 120 minute time integration using a  time step of $30~s$.  
  Figure \ref{fig:MtnRidge} displays the vertical velocity at 120 simulated minutes from which it can be seen to be in close agreement with the results in \cite{Klemp2015}.  

\subsubsection{Circular Mountain Case}
\begin{figure}[ht]%
	\centering
	\includegraphics[width=0.75\linewidth]{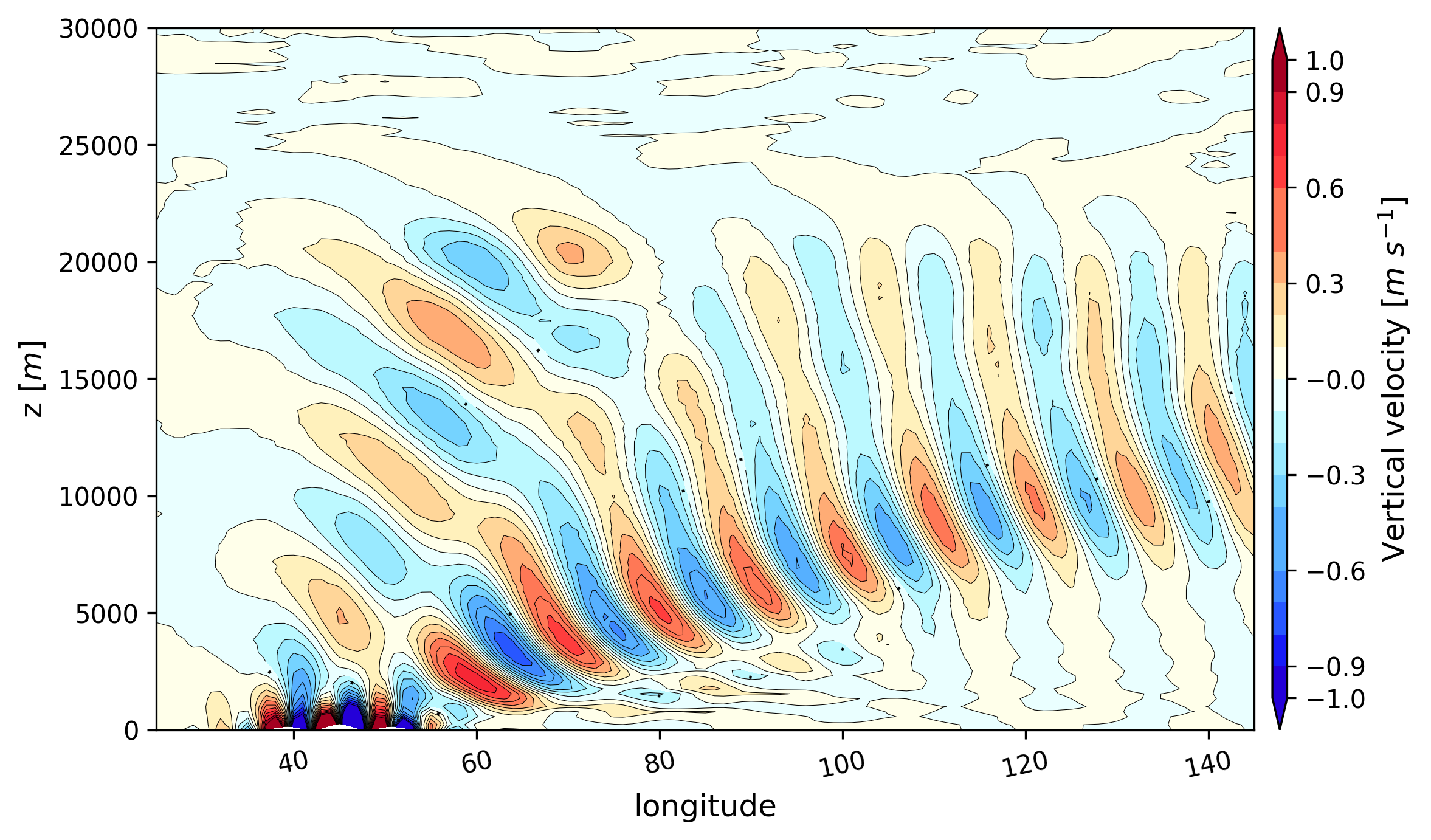}
	\caption{Vertical velocity along equator after 120 minutes simulation of the circular mountain case.  The vertical velocity contour interval is $0.1$ $m/s$.  }
	\label{fig:circular_mtwave}
\end{figure}

For this case, we consider uniform flow over a circular mountain of height $250~m$. Using third-order spectral element polynomials, we set our model top at $30~km$ with $118$ vertical levels ($39$ elements) for approximately $250~m$ spacing and consider a horizontal grid spacing of 0.55 degree (approx. $360~m$) using $56$ elements per face of the cubed sphere which puts us in the nonhydrostatic regime. Figure \ref{fig:circular_mtwave} shows the modeled vertical velocity along the equator after 120 min integration using the exponential Euler method with a $30~s$ time step.

It is clear that the exponential integrator is able to capture the gravity waves consisting of both the larger-scale hydrostatic waves (propagating vertically) and the smaller-scale nonhydrostatic waves (propagating downstream). Comparing this modeled solution (Figure \ref{fig:circular_mtwave}) with the results from Fig. 4/5 of \cite{Klemp2015} indicate that the simulations with the exponential scheme(s) reproduce all the essential features of this case.

\subsection{Baroclinic Instability}\label{subsec:baroclinic}
A commonly used idealized test for model developers is baroclinic instability \cite{Ullrich2014}.  The test consists of making a very small amplitude perturbation on an unstable zonal jet to an initially balanced state such that it triggers the development (over several days) of a baroclinic wave in the northern hemisphere.  The instability appears near day five and reaches maturity day nine in the simulation and continues to grow over a 15-day simulation.  Here, we will employ \textit{EXP2} and assess how it responds to a controlled baroclinic instability.

\begin{figure}[hbt]%
	\centering
	\subfigure[][Surface pressure (day 8)]{%
		\label{fig:bwave_sfc_press_day8}
		\includegraphics[width=0.49\linewidth]{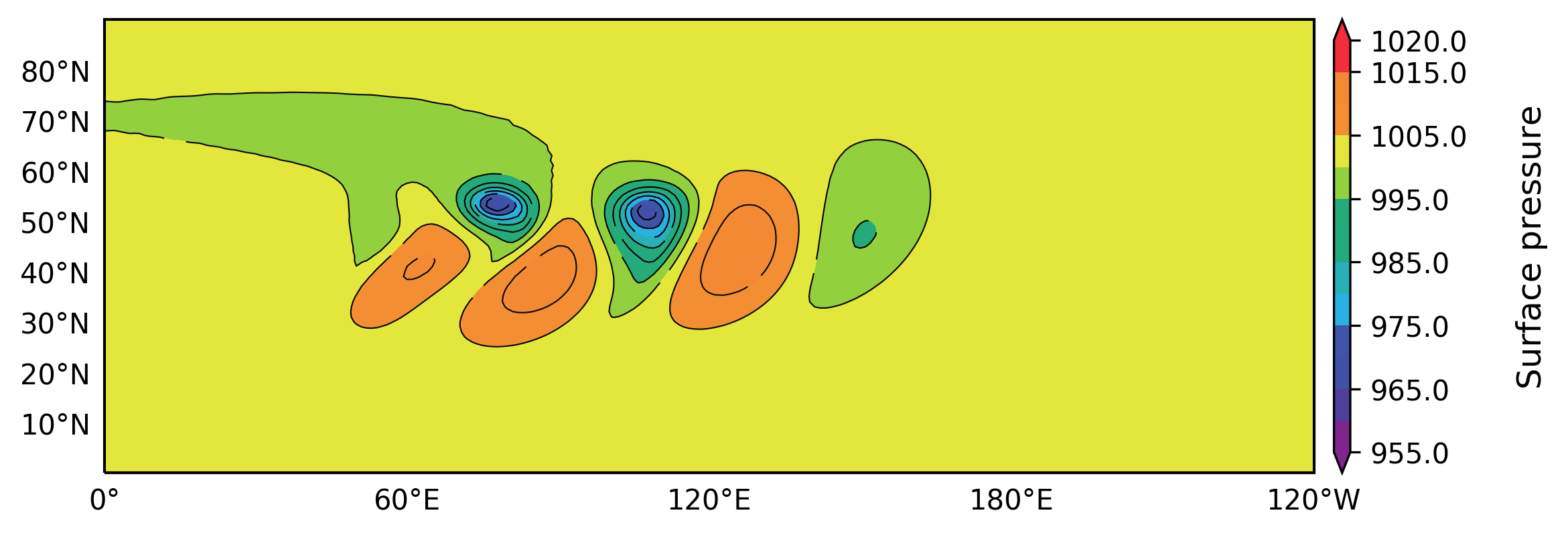}	}%
	\subfigure[][Surface pressure (day 10)]{%
	\label{fig:bwave_sfc_press_day10}
	\includegraphics[width=0.49\linewidth]{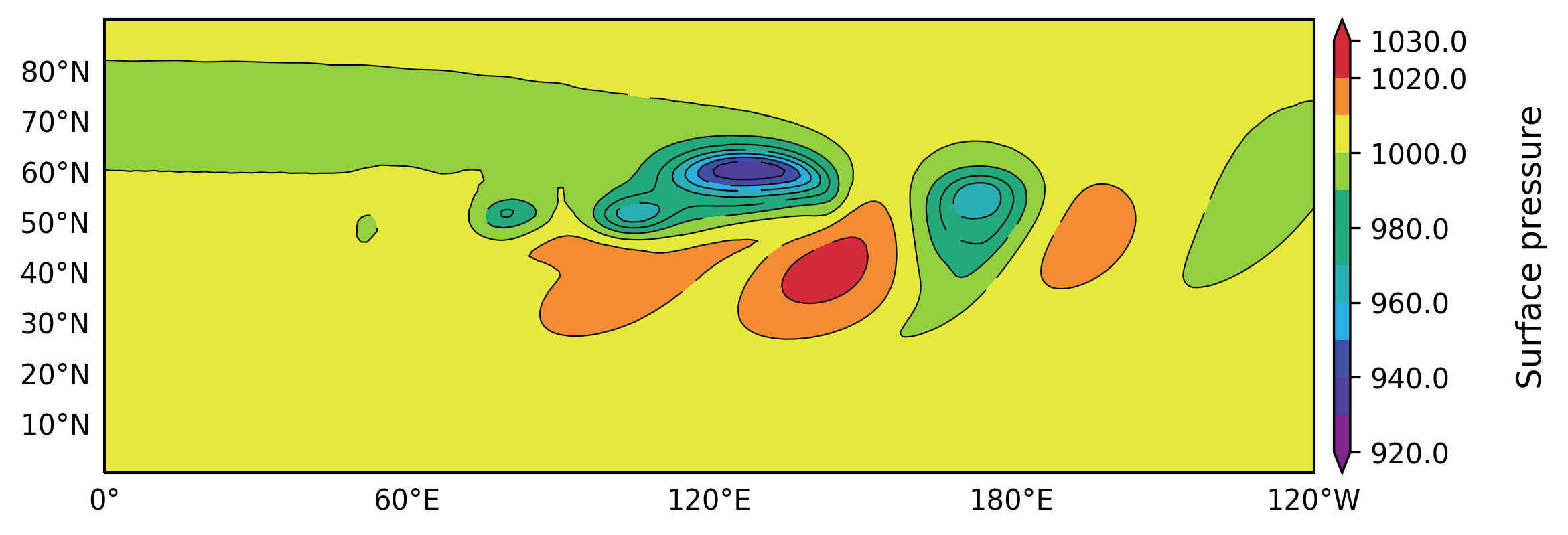}	}\\%
	\subfigure[][$850~hPa$ relative vorticity (day 8)]{%
		\label{fig:bwave_vort_day8}
		\includegraphics[width=0.49\linewidth]{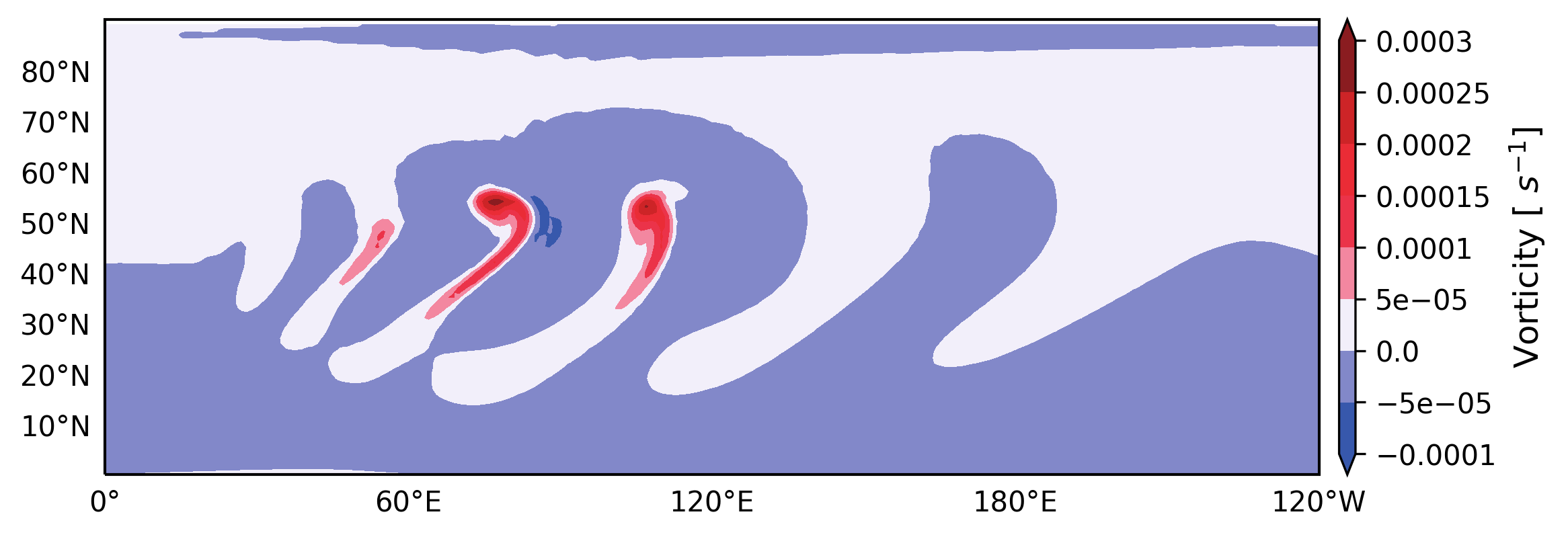}}
		\subfigure[][$850$ hPa relative vorticity (day 10)]{%
		\label{fig:bwave_vortz_day10}
		\includegraphics[width=0.49\linewidth]{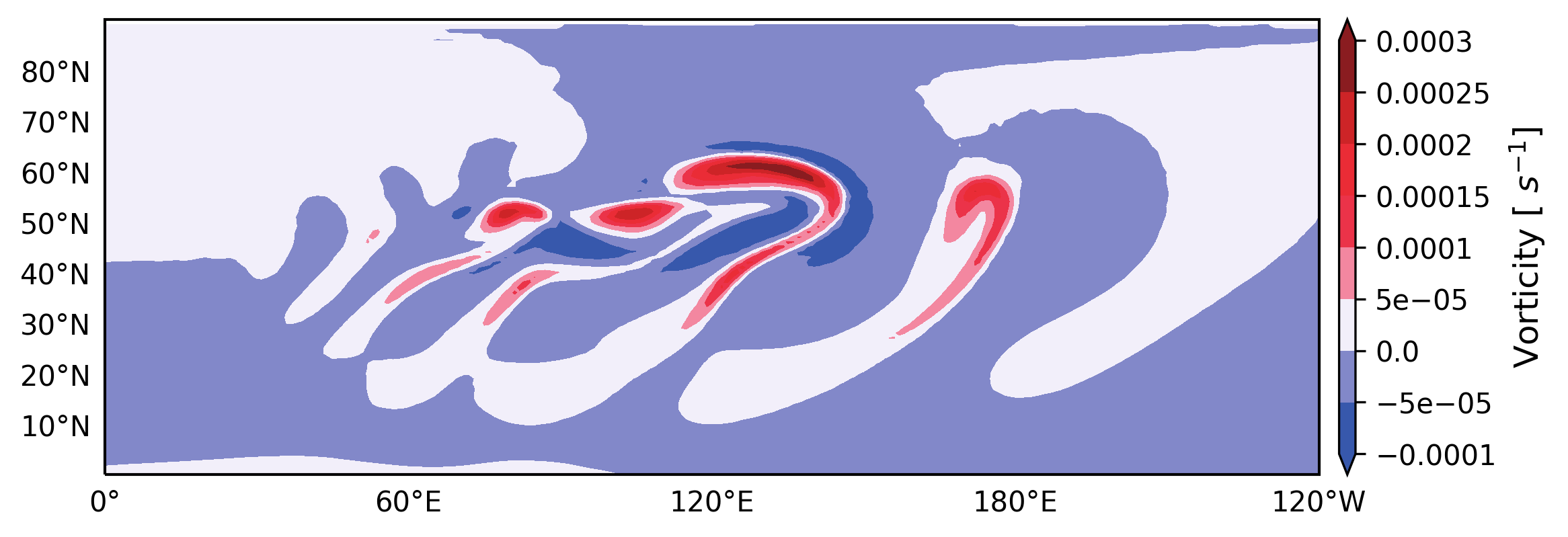}}\\			
	\caption[  ]{Surface pressure (top row) and $850$ hPa relative vorticity (bottom row) shown at days 8 (left column) and 10 (right column) of the baroclinic instability simulation.  }
	\label{fig:bwave}
\end{figure}

\begin{figure}[ht]%
	\centering
	\includegraphics[width=0.4\linewidth]{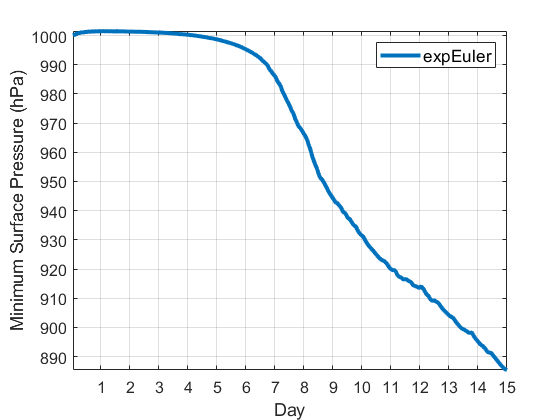}
	\caption{Time-series plot of the minimum surface pressure over the 15 day baroclinic instability simulation}
	\label{fig:bwave_minsfcpress}
\end{figure}

Our numerical setup for this test case uses third-order spectral element polynomials, 32 elements per edge of the cubed sphere, and 31 vertical elements with a model top of $30~km$. This discretization corresponds to $1^\circ$ (approx. $111~km$) horizontal grid spacing and near $1.03~km$ in the vertical. The exponential Euler method was used to propagate the solution in time using a time step of $360~s$. Figure \ref{fig:bwave} displays the surface pressure and the 850 hPa relative vorticity at days 8 and 10 of the simulation from which the baroclinic instability can be seen to be well underway and in agreement with the results in \cite{Ullrich2014}. A plot of the minimum surface pressure over the 15-day simulation is provided in Figure \ref{fig:bwave_minsfcpress};  showing the evolution of the minimum surface pressure is within the packet of solutions from the suite of dynamical cores tested in \cite{Ullrich2014}.

 \subsection{Real-Data Forecast}\label{subsec:realdata}
 We conclude our numerical experiments with a real-data simulation that confirms exponential integrators are suitable for NWP applications.  
The simulations presented here specified a model top of 51 km (with an upper boundary absorbing layer of depth $10 ~km$) using 32 elements per face of the cubed sphere, 3rd order polynomials, and 64 vertical levels (21 elements) which corresponds to $1^\circ$ ($111~km$) horizontal grid spacing and approximately $0.8~km$ in the vertical.  Using initial conditions that were interpolated to the NEPTUNE grid from NOAA Environmental Modeling System (NEMS) / (FV3-GFS) data on 0000 UTC 1 January 2019, we ran a 96-hour forecast using \textit{EXP2} with a 30 second time step using the GFS physics \cite{CCPP}.  To show that we obtain a reasonable forecast, we ran a second simulation using the horizontally-explicit-vertically-implicit (HEVI) implementation \cite{gardneretal2018} of the 2nd-order additive Runge-Kutta method \textit{ars232} \cite{ASCHER1997151}.  It is important to note that the time step size was chosen to be compatible with the physics as tuned for the \textit{ars232}, and larger time steps can be accommodated by sub-stepping the physics (see conclusion).

\begin{figure}[h!]%
 	\centering	
 	\subfigure[][\textit{EXP2}: 0000 UTC 2 January 2019 ]{%
 		\label{fig:realdata_24h_exp}
 		\includegraphics[width=0.45\linewidth]{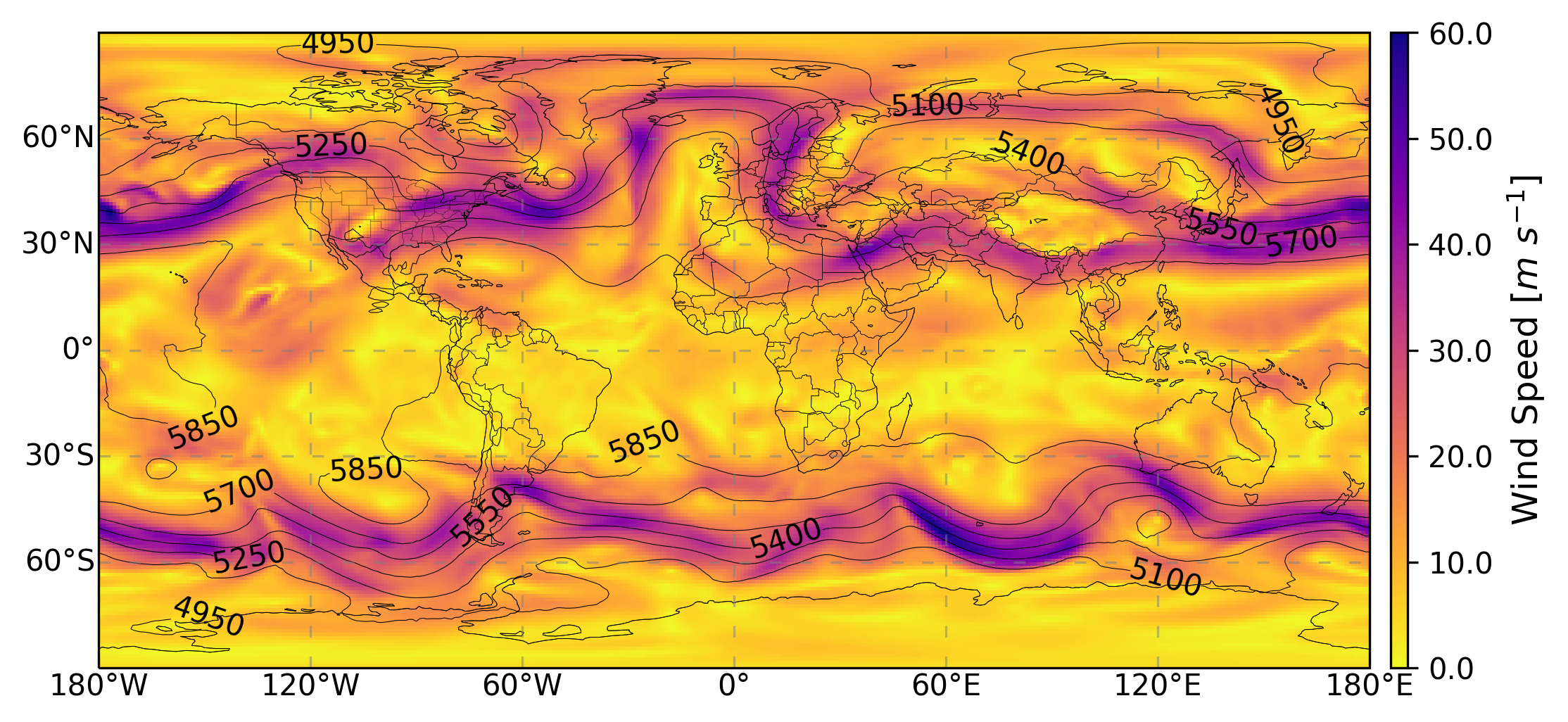}	}%
 	\hspace{8pt}%
 	\subfigure[][\textit{ars232}:  0000 UTC 2 January 2019]{%
 		\label{fig:realdata_24h_ars232}
 		\includegraphics[width=0.45\linewidth]{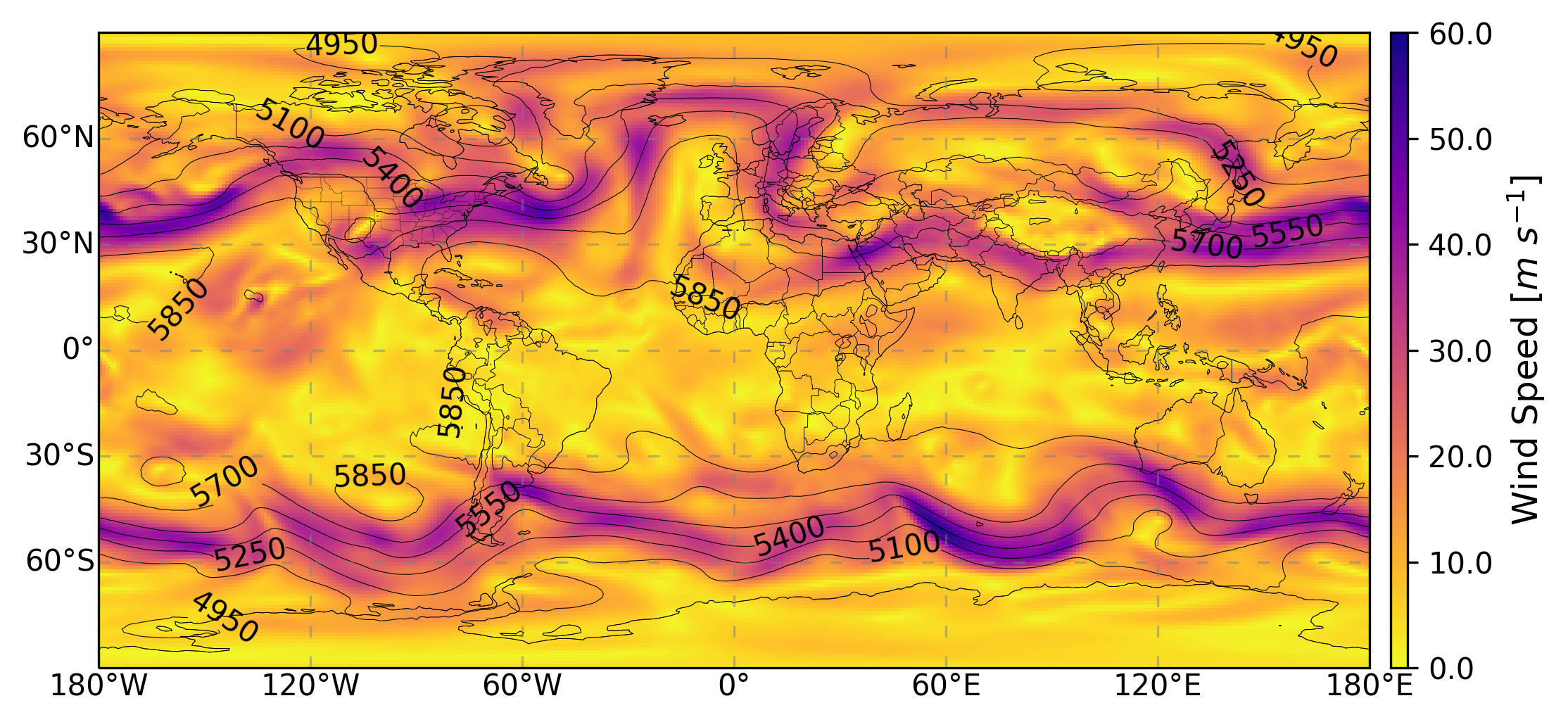}}\\	
 	\subfigure[][\textit{EXP2}: 0000 UTC 3 January 2019 ]{%
 		\label{fig:realdata_48h_exp}
 		\includegraphics[width=0.45\linewidth]{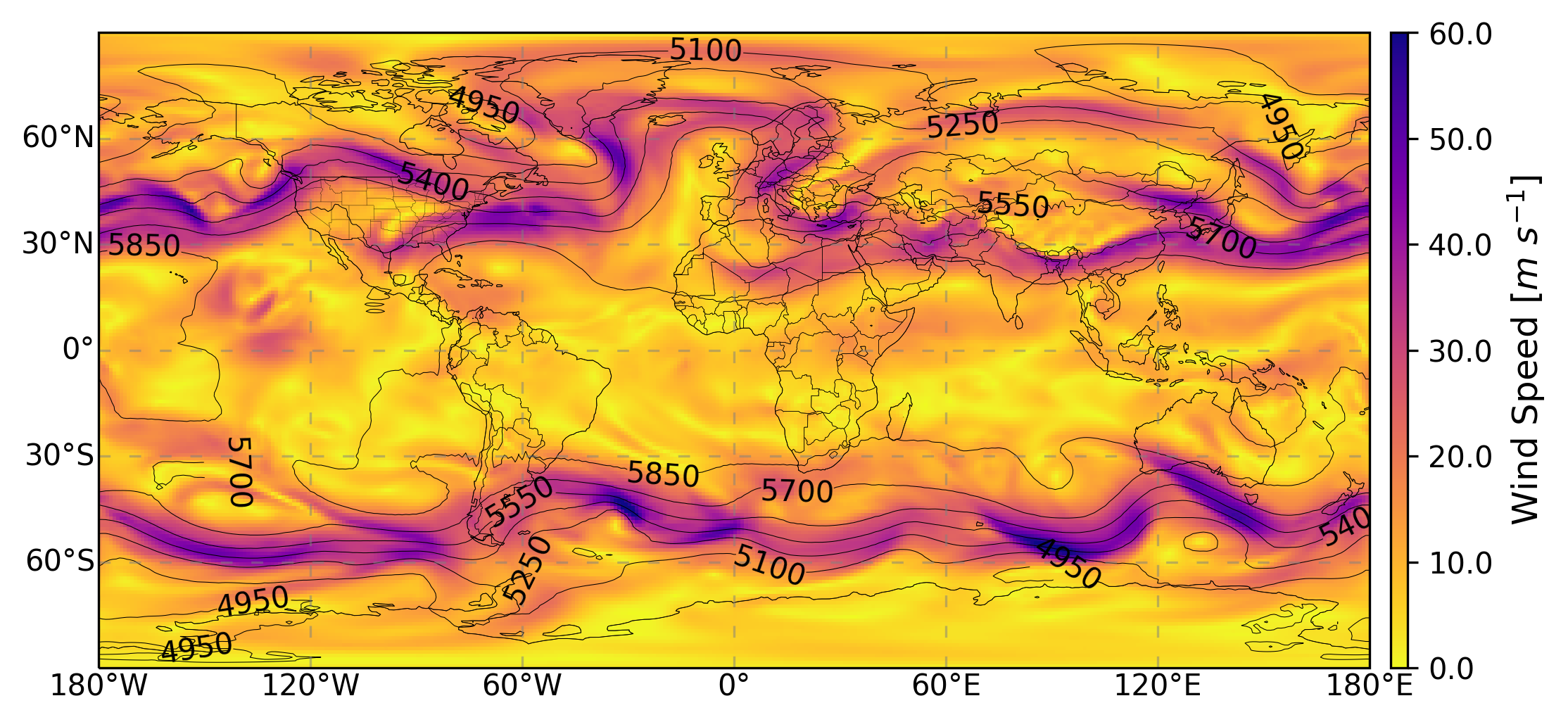}}%
 	\hspace{8pt}%
 	\subfigure[][\textit{ars232}:  0000 UTC 3 January 2019]{%
 		\label{fig:realdata_48h_ars232}
 		\includegraphics[width=0.45\linewidth]{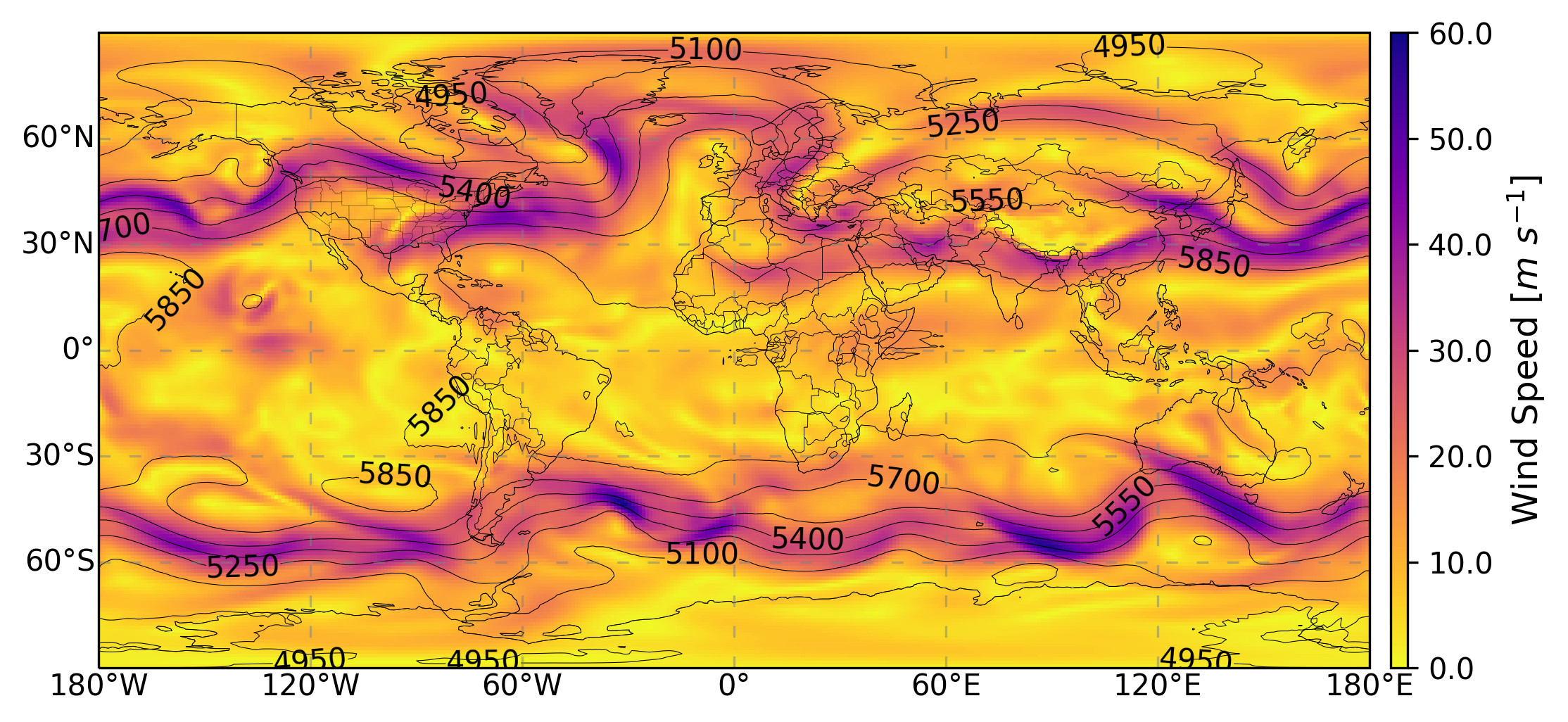}}\\
 	\subfigure[][\textit{EXP2}: 0000 UTC 4 January 2019 ]{%
 		\label{fig:realdata_72h_exp}
 		\includegraphics[width=0.45\linewidth]{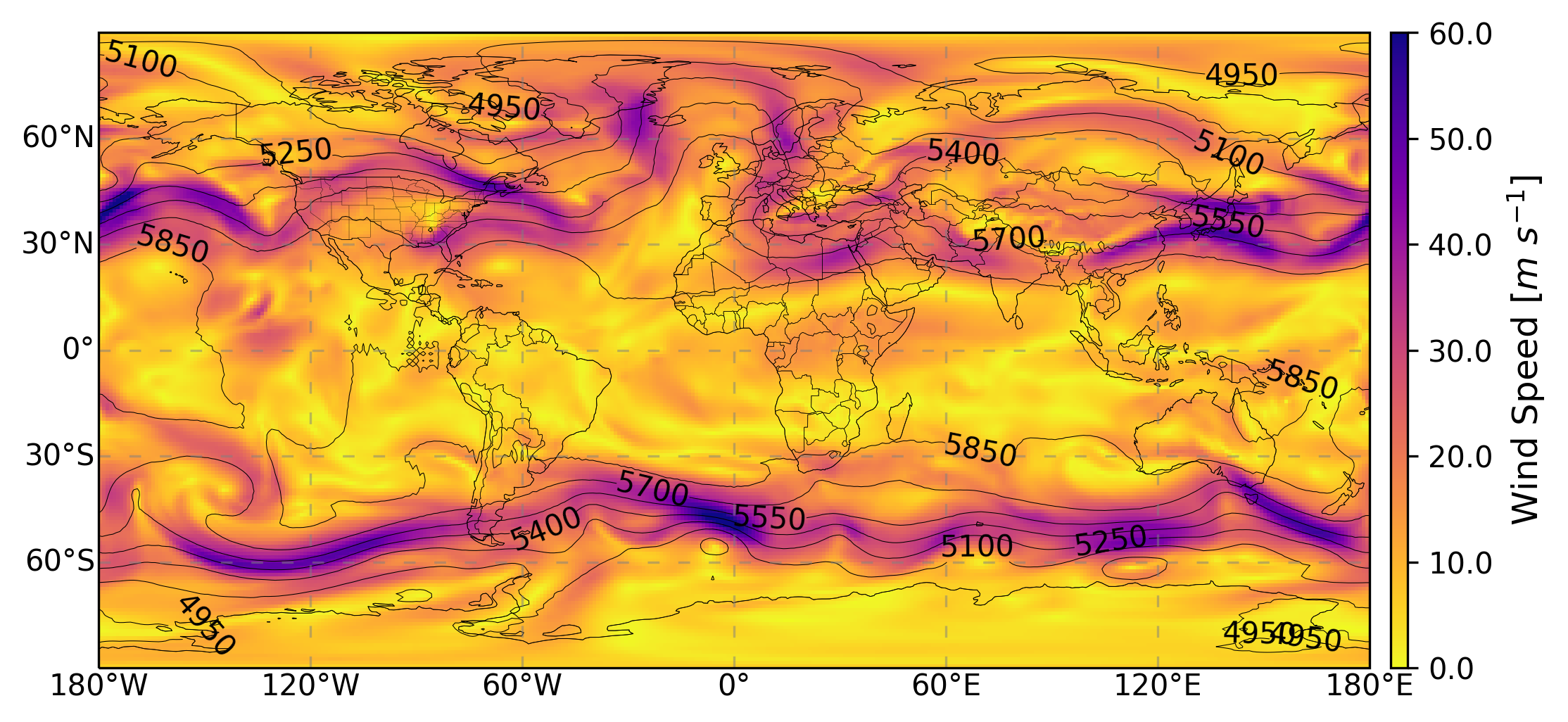}}%
 	\hspace{8pt}%
 	\subfigure[][\textit{ars232}:  0000 UTC 4 January 2019]{%
 		\label{fig:realdata_72h_ars232}
	\includegraphics[width=0.45\linewidth]{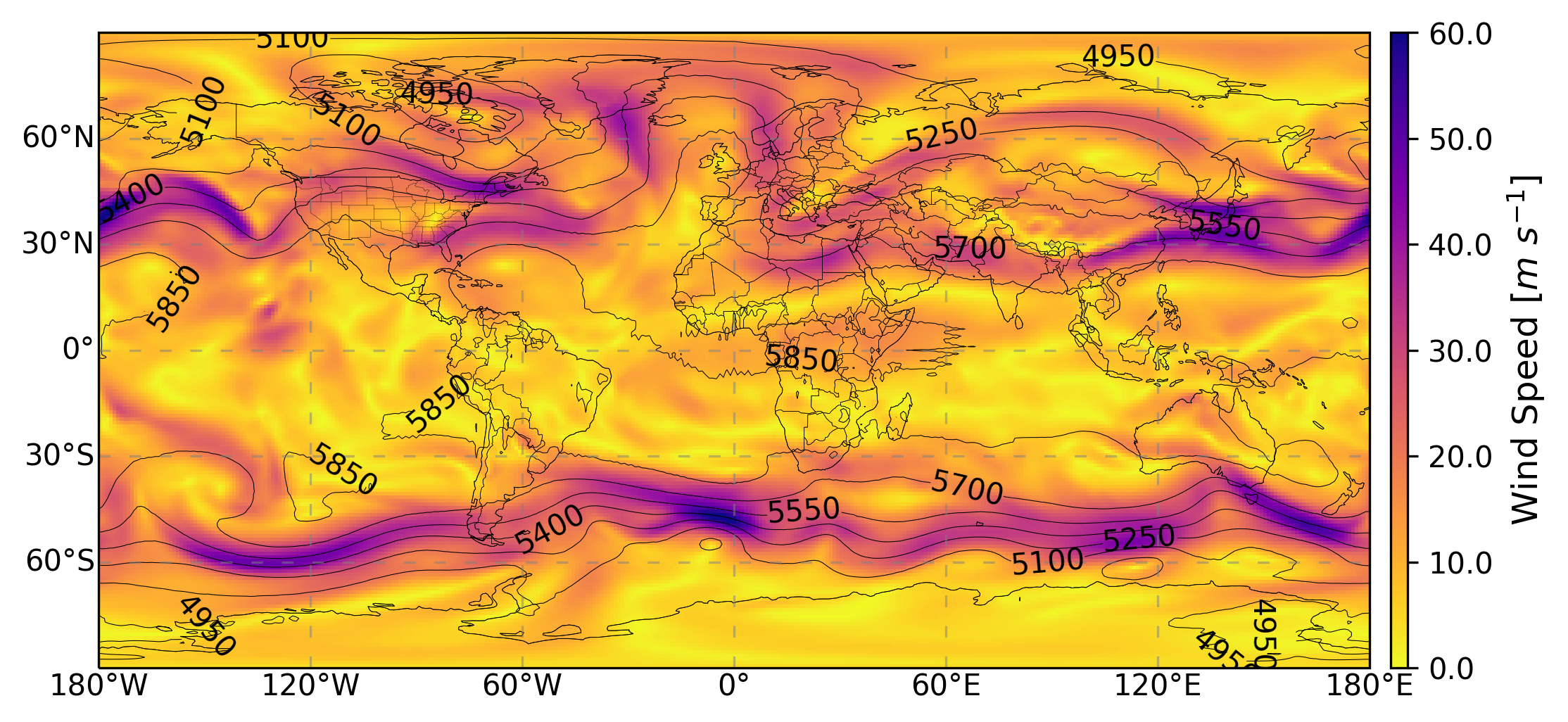}}\\
 	\subfigure[][\textit{EXP2}: 0000 UTC 5 January 2019]{%
 		\label{fig:realdata_96h_exp}
 		\includegraphics[width=0.45\linewidth]{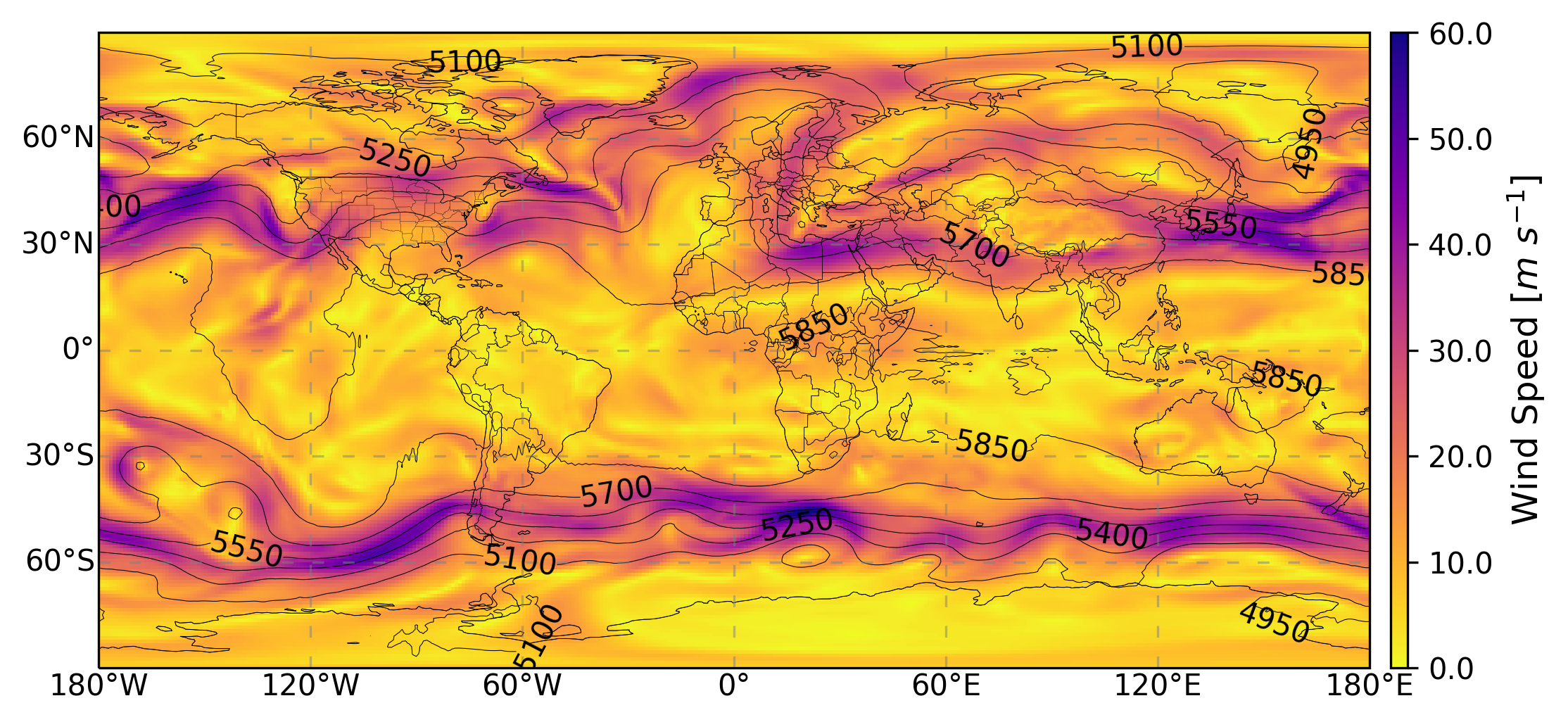}}%
 	\hspace{8pt}%
 	\subfigure[][\textit{ars232}:  0000 UTC 5 January 2019]{%
 		\label{fig:realdata_96h_ars232}
	\includegraphics[width=0.45\linewidth]{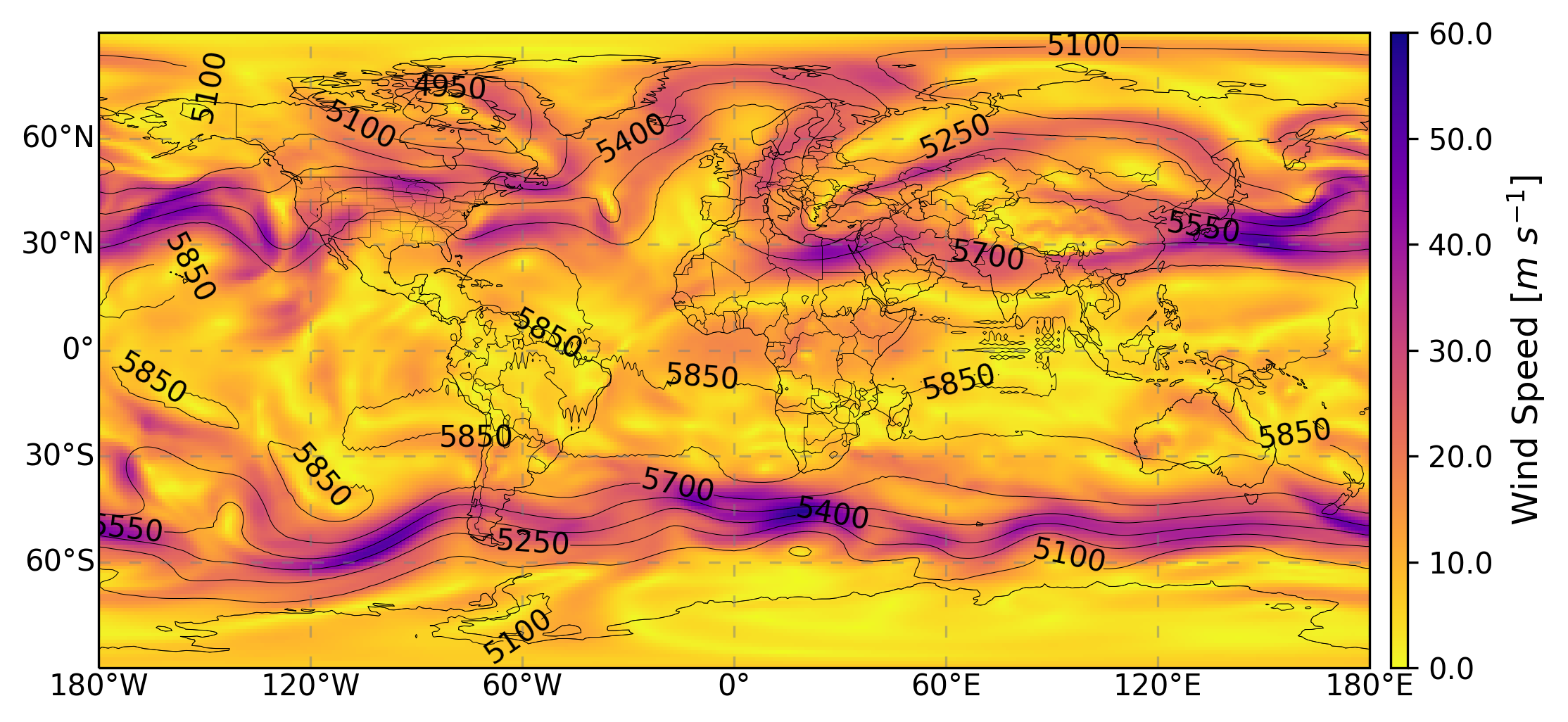}}\\
 	\caption{Real data simulation initialized on 0000 UTC 1 January 2019. The shading denotes the wind speed and contours represent the geometric height at 500 $hPa$ with a contour interval of 150 $m$ }\label{fig:realdata}
 \end{figure}

Figure \ref{fig:realdata} displays the wind speed and geometric heights at 24 hour increments of the forecast for \textit{EXP2} and \textit{ars232}. It is clear that \textit{EXP2} captures the large-scale flow of the mid-latitude jets, and upon comparison with \textit{ars232}, we see the same large-scale features, including a large-scale omega block in the North Atlantic (for further discussion of the synoptic setup see \cite{Yessimbet}). Thus, exponential methods are able to run with full physics and produce realistic forecasts.

\section{Summary and conclusions}
Using the Navy's next generation numerical weather prediction system, NEPTUNE, we demonstrated that exponential methods are suitable time integrators for 3D compressible deep-atmosphere nonhydrostatic models.  Taking advantage of the stability of exponential integrators, we showed that large time steps can be taken while still resolving both the large- and small- scale dynamics.  The idealized test case simulations are comparable to those in the literature and provide the necessary confirmation that exponential methods are able to accurately simulate both hydrostatic and nonhydrostatic processes.  Additionally, the real data experiment indicates the ability to handle the coupling of physics with dynamics and produce forecasts that align closely with the \textit{ars232} IMEX time integrator at the same time step size.

It is of future work to conduct comprehensive comparisons with other time integrators and sub-stepping of the physics. Preliminary results suggest that \textit{EXP2} can be competitive with IMEX and split-explicit time integration schemes.  We do note that the methods of order four presented here are deemed too costly due to the fact that they require two Krylov projections per time step.  However, \cite{GAUDREAULT2021} recently developed exponential multistep methods up to order six that require only a single Krylov projection per time-step.  These methods are of great interest as the high accuracy complements the spectral element spatial discretization with a computational cost comparable to \textit{EXP2}.

\section*{Acknowledgments}
This work was supported by an American Society for Engineering Education Postdoctoral Fellowship and the Office of Naval Research 6.1-IDPI (Program Element 0601153N).

\appendix
\section{Analysis of the absorbing upper boundary layer for exponential methods}\label{sec:appendixAnalysis}
Our analysis is based on partitioning
\begin{equation}
\frac{\partial \mathbf{u}}{\partial t}=f(t,\mathbf{u}(t)), \quad \mathbf{u}(t_0)=\mathbf{u}_0, \quad \mathbf{u}(t)\in \mathbb{R}^n \label{eqn:Appendix_IVP}
\end{equation}
as 
\begin{equation}\label{eqn:partRHS}
\frac{\partial \mathbf{u}}{\partial t} = \underbrace{\left\{f(t,\mathbf{u})-L\mathbf{u}\right\}}_{N(t,\mathbf{u})} +[L\mathbf{u}]
\end{equation}
where $L$ is a linear operator that contains the terms responsible for the fast acoustic and gravity waves.  The exponential Euler scheme for the partitioned problem \eqref{eqn:partRHS} is given by
\begin{equation}
\label{eqn:partitionedExpEuler}
\mathbf{u}(t_n+\dt)\approx \mathbf{u}(t_n)+\varphi_1(\dt L)\dt (L\mathbf{u}(t_n)+N(t_n,\mathbf{u}(t_n))).
\end{equation}
For ease in presenting our analysis, we re-rewrite the adjustment \eqref{eqn:EulerAdjustment} as a pre-time step adjustment:
\begin{equation}\label{eqn:part_euler_adjustment}
\left\{\begin{array}{l}
\widehat{\mathbf{u}}^n =\mathbf{Q} \mathbf{u}^n \\
\mathbf{u}^{n+1}=\widehat{\mathbf{u}}^n +\dt \varphi_1(\dt L)(N(t_n,\widehat{\mathbf{u}}^n) +L\widehat{\mathbf{u}}^n)
\end{array}\right.,
\end{equation}	
where $\mathbf{Q}=\textrm{diag}\left[1,1,1,(1+\dt R_w)^{-1},1\right]$.

We note that method \eqref{eqn:partitionedExpEuler} is first-order accurate.   Our analysis is based on the linear \textit{IMEX} formulation \cite{giraldoetal2013} which considers a splitting of the density $\rho(\mathbf{x},t)=\rho_0(\mathbf{x})+\rho'(\mathbf{x},t)$, potential temperature $\theta(\mathbf{x},t)=\theta_0(\mathbf{x})+\theta'(\mathbf{x},t)$, and pressure $P(\mathbf{x},t)=P_0(\mathbf{x})+P'(\mathbf{x},t)$, where the reference values are in hydrostatic balance.  We then consider the partitioning \eqref{eqn:partRHS} with the linear operator defined as
\begin{equation}
\label{eqn:linOp}
L(\mathbf{u})=-\begin{pmatrix}w\dfrac{d\rho_0}{dr}+\rho_0\dfrac{\partial w}{\partial r}\\0 \\0 \\ \dfrac{1}{\rho_0}\dfrac{\partial P'}{\partial r}+g\dfrac{\rho'}{\rho_0}\\ w\dfrac{d\theta_0}{dr}\end{pmatrix}, \end{equation}
where $w$ is the vertical velocity, $r$ is the radial direction from the center of the sphere, $g$ is gravity, and the pressure is defined via the linearization of the equation of state \begin{equation}
\label{eqn:pressure}
P=p_{ref}\left(\frac{\rho R\theta}{p_{ref}}\right)^{\gamma},
\end{equation} as
\begin{equation}
\label{eqn:pressure_linearized}
P'=G_0\rho'+H_0\theta'
\end{equation}
with
\begin{eqnarray}
G_0 &=& \frac{\gamma P_0}{\rho_0} \label{eqn:G0}\\
H_0 &=& \frac{\gamma P_0}{\theta_0} \label{eqn:H0},
\end{eqnarray} 
and where $R=c_p-c_v$ is the specific gas constant, $c_p$ and $c_v$ are mass specific heats at constant pressure and volume, and $\gamma = c_p /c_v$.

We begin with the derivation of the dominant damping terms for the first order exponential scheme  (\ref{eqn:partitionedExpEuler}) and then show that we obtain the same terms for any higher-order exponential Runge-Kutta scheme.  At time $t_n$, we make an adjustment to our solution $\mathbf{u}^n$:
\begin{equation}\label{eqn:q_adjust}
\widehat{\mathbf{u}}^n =\mathbf{Q} \mathbf{u}^n=\vect{\rho'^n\\ u^n \\ v^n \\ (1+\dt R_w)^{-1} w^n\\ \theta'^n}.
\end{equation}
Using the approximation $(1+z)^{-1}\approx 1-z$, our adjustment (\ref{eqn:q_adjust}) can be expressed as
\begin{equation}\label{eqn:q_adjust2}
\widehat{\mathbf{u}}^n = \mathbf{u}^n - \dt \vect{0\\ 0\\ 0 \\ R_w w^n\\ 0}.
\end{equation}
Inserting (\ref{eqn:q_adjust2}) into  (\ref{eqn:part_euler_adjustment}) for $\mathbf{u}^n$ we obtain (via linearity of $L$ and some re-arranging)
\begin{equation}\label{eqn:expEuler_3}
\mathbf{u}^{n+1} = \mathbf{u}^n + \dt \varphi_1(\dt L)(L\mathbf{u}^n +N(t_n,\widehat{\mathbf{u}}^n)) \underbrace{-  \dt \vect{0\\ 0\\ 0 \\ R_w w^n\\ 0} - \dt^2\varphi_1(\dt L)\left(L\vect{0\\ 0\\ 0 \\ R_w w^n\\ 0}\right)}_{\textrm{ damping terms}}.
\end{equation}
The last two terms on the right of (\ref{eqn:expEuler_3}) are responsible for the absorption of gravity waves and are, therefore, our primary focus. The first of these terms produces the Raleigh damping term in the vertical momentum equation.  Using the Taylor series expansion of 
$$\varphi_1(z)=\frac{e^z-1}{z}= 1+ \frac{z}{2!}+\frac{z^2}{3!}+\frac{z^3}{4!}+\dots$$
we can express the last term on the right-hand-side of (\ref{eqn:expEuler_3}) as
\begin{eqnarray}\label{eqn:phi1expansion}
\dt^2\varphi_1(\dt L)\left(L\vect{0\\ 0\\ 0 \\ R_w w^n\\ 0}\right) &=& \dt^2 L\vect{0\\ 0\\ 0 \\ R_w w^n\\ 0} + \frac{\dt^3}{2!}L^2\vect{0\\ 0\\ 0 \\ R_w w^n\\ 0}+\dots,
\end{eqnarray}
for which the following damping terms are obtained \\
$\rho'$:
\begin{eqnarray}
-\dt^2\left( \frac{d\rho_0}{dr} R_w w^n + \rho_0 \dfrac{\partial}{\partial r}\left(R_w w^n\right) \right) +\mathcal{O}(\dt^3)\label{eqn:rhoDamping}
\end{eqnarray}

\noindent $w$:
\begin{eqnarray}
&&\dt^3 \left[\left(\frac{1}{\rho_0} \frac{d}{dr}\left(G_0 \frac{d\rho}{dr}+H_0\frac{d\theta_0}{dr} \right)+\frac{g}{\rho_0}\frac{d\rho_0}{dr}\right)R_w w^n 
+\left(\frac{1}{\rho_0}\left(G_0\frac{\rho_0}{dr}+H_0\frac{d\theta_0}{dr}\right)+\frac{1}{\rho_0}\frac{d}{dr}(G_0\rho_0)+g\right)\frac{\partial}{\partial r}\left(R_w w^n\right) \right. \nonumber \\
&&\qquad +\left. G_0\rho_0 \frac{\partial^2}{\partial r^2}\left(R_w w^n\right)\right]+\mathcal{O}(\dt^4) \label{eqn:wDamping}
\end{eqnarray}

\noindent $\theta'$:
\begin{eqnarray}
-\dt^2 \frac{d\theta_0}{dr} R_w w^n+\mathcal{O}(\dt^3) \label{eqn:thetaDamping}
\end{eqnarray}
Upon inserting \eqref{eqn:wDamping} into \eqref{eqn:expEuler_3}, we can express our vertical momentum equation as
\begin{eqnarray}
\frac{\partial w}{\partial t} \approx \mathbf{P}\left(\frac{\mathbf{u}^{n+1}-\mathbf{u}^n}{\dt}\right) &=& \mathbf{P}\left(\varphi_1(\dt L)(L\mathbf{u}^n +N(\widehat{\mathbf{u}}^n))\right)-R_w w^n - \dt^2\left[\left(\frac{1}{\rho_0} \frac{d}{dr}\left(G_0 \frac{d\rho}{dr}+H_0\frac{d\theta_0}{dr} \right)\right.\right.\left. +\frac{g}{\rho_0}\frac{d\rho_0}{dr}\right)R_w w^n \nonumber
\\ 
& & \qquad  \left. 
+\left(\frac{1}{\rho_0}\left(G_0\frac{\rho_0}{dr}+H_0\frac{d\theta_0}{dr}\right)+\frac{1}{\rho_0}\frac{d}{dr}(G_0\rho_0)+g\right)\frac{\partial}{\partial r}\left(R_w w^n\right) \right. \nonumber \\
&&\qquad +\left. G_0\rho_0 \frac{\partial^2}{\partial r^2}\left(R_w w^n\right)\right]+\mathcal{O}(\dt^3),\label{eqn:damping_vme}
\end{eqnarray}
where $\mathbf{P}$ is a 5 by 5 projection matrix with $\mathbf{P}_{44}=1$ and $\mathbf{P}_{ij}=0$.  Thus the adjustment \eqref{eqn:EulerAdjustment} to the vertical velocity produces not only the Raleigh damping term $R_w w^n$, but also second order terms (in time) that includes a vertical diffusion-like term $G_0\rho_0 \frac{\partial^2}{\partial r^2}\left(R_w w^n\right)$ (cf. \cite{klempetal2008}).  The same damping terms are obtained for higher-order Runge-Kutta methods as they can be viewed as a perturbation of the exponential Euler scheme (\ref{eqn:partitionedExpEuler}).
%

\bibliographystyle{amsplain}
\bibliography{bib}
\end{document}